\documentclass[aps,twocolumn,showkeys,showpacs,superscriptaddress,preprintnumbers]{revtex4}
\usepackage[english]{babel}
\usepackage{amsfonts, amsmath}
\usepackage{graphicx}
\usepackage{color}

\newcommand{\trn}{^{\scriptscriptstyle {\rm T}}}
\begin{document}

\title{Adaptive synchronization in delay-coupled networks of Stuart-Landau
oscillators}

\author{Anton~Selivanov}
\affiliation{Department of Theoretical Cybernetics, Saint-Petersburg State University, Saint-Petersburg, Russia}
\author{Judith~Lehnert}
\affiliation{Institut f\"ur Theoretische Physik, Technische Universit\"at Berlin, Hardenbergstr. 36, 10623 Berlin, Germany}
\author{Thomas~Dahms}
\affiliation{Institut f\"ur Theoretische Physik, Technische Universit\"at Berlin, Hardenbergstr. 36, 10623 Berlin, Germany}
\author{Philipp~H\"ovel}
\affiliation{Institut f\"ur Theoretische Physik, Technische Universit\"at Berlin, Hardenbergstr. 36, 10623 Berlin, Germany}
\author{Alexander~Fradkov}
\affiliation{Department of Theoretical Cybernetics, Saint-Petersburg State University, Saint-Petersburg, Russia}
\author{Eckehard~Sch\"oll}
\email[corresponding author: ]{schoell@physik.tu-berlin.de}
\affiliation{Institut f\"ur Theoretische Physik, Technische Universit\"at Berlin, Hardenbergstr. 36, 10623 Berlin, Germany}

\date{\today}
%\preprint{SEL11, draft from \today}

\begin{abstract}
We consider networks of delay-coupled Stuart-Landau oscillators. In these systems,
%  As it was shown in (C.-U. Choe, T. Dahms, P. H\"ovel, and E. Sch\"oll, Phys. Rev. E 81, 025205(R), 2010)
the coupling phase has been found to be a crucial control
parameter. By proper choice of this parameter one can switch between
different synchronous oscillatory states of the network. Applying the
speed-gradient method,
% (A. L. Fradkov, Cybernetical Physics: From Control of Chaos to Quantum Control (Springer, Heidelberg, Germany, 2007))
we derive an adaptive algorithm for an automatic adjustment of the
coupling phase such that a desired state can be selected from an
otherwise multistable regime. We propose goal functions based on both
the difference of the oscillators and a generalized order parameter
and demonstrate that the speed-gradient method allows one to find
appropriate coupling phases with which different states of
synchronization, e.g., in-phase oscillation, splay or various
cluster states, can be selected.
\end{abstract}

% Synchronization, nonlinear dynamics, 05.45.Xt
% Complex systems, 89.75.-k
% Control theory in mathematical physics, 02.30.Yy
% Delay equations, in function theory, 02.30.Ks

\maketitle

\section{Introduction}
The ability to control nonlinear dynamical systems has brought up a
wide interdisciplinary area of research that has evolved rapidly
in the last decades \cite{SCH07}. In particular, noninvasive control
schemes based on time-delayed feedback \cite{PYR92,SOC94,PYR06} have been
studied and applied to various systems ranging from biological and
chemical applications to physics and engineering in both theoretical
and experimental works \cite{POP06,SCH06a,ZHA08,ORO10,DAH10,SCH09,SCH09a,FLU11}. 
Here we propose to use adaptive control schemes based on optimizations of 
cost or goal functions \cite{FRA79,FRA05b,FRA07} to find appropriate control parameters. 
Besides isolated systems, control of dynamics in
spatio-temporal systems and on networks has recently gained much
interest \cite{KEH09,HOE09,FLU10b,OME11,BRO11}. The existence and control of
cluster states was studied by Choe {\em et al.}
\cite{CHO09,CHO11} in networks of Stuart-Landau oscillators. This
Stuart-Landau system arises naturally as a generic expansion near a
Hopf bifurcation and is therefore often used as a paradigm for
oscillators.  The complex coupling constant that arises
from the complex state variables in networks of Stuart-Landau
oscillators consists of an amplitude and a phase. Similar coupling
phases arise naturally in systems with all-optical coupling
\cite{SCH06a,FLU07}. Such phase-dependent couplings have also been
shown to be important in overcoming the odd-number limitation of
time-delay feedback control \cite{FIE07,SCH11} and in anticipating
chaos synchronization \cite{PYR08}. Furthermore, it was shown in
\cite{CHO09,CHO11} that the value of the coupling phase is a crucial
control parameter in these systems, and by adjusting this phase one can
deliberately switch between different synchronous oscillatory states
of the network. In order to find an appropriate value of the coupling
phase one could solve a nonlinear equation that involves the system
parameters. However, in practice the exact values of the system parameters
are unknown, and analytical conditions can be derived only for special
values of the complex phase. An efficient way to avoid these limitations 
and find optimal values of the coupling phase is the use of adaptive
control.

In this paper, we present an adaptive synchronization algorithm for
delay-coupled networks of Stuart-Landau oscillators. To find an
adequate coupling phase we apply the speed-gradient method
\cite{FRA07}, which was used previously in various nonlinear control
problems, yet not for the control of dynamics in delay-coupled networks. 
By taking an appropriate goal function we derive an equation
for the automatic adjustment of the coupling phase such that the goal
function is minimized. At the same time the coupling phase converges
to the theoretically predicted value. Our goal function is based on
the Kuramoto order parameter and is able to distinguish the different
states of synchrony in the Stuart-Landau networks irrespectively of
the numbering of the nodes.

This paper is organized as follows. After this introduction, we
describe the model system in
Sec.~\ref{sec:model}. Section~\ref{sec:SG} introduces the
speed-gradient method and its application using the coupling phase in
networks of Stuart-Landau oscillators. We present the main results for
the control of in-phase synchronization in
Sec.~\ref{sec:in-phase-synchr}, and for cluster and splay states in
Sec.~\ref{sec:splay-cluster-states}. Finally, Sec.~\ref{sec:conclusions}
contains some conlclusions.

\section{Model equation}\label{sec:model}
% \section{Problem statement}
Consider a network of $N$ delay-coupled oscillators
\begin{equation}
    \label{mainEq}
    \dot{z}_j(t)=f[z_j(t)]+Ke^{i\beta}\sum_{n=1}^N a_{jn}[z_n(t-\tau)-z_j(t)]
% , \quad j=1\ldots N,
\end{equation}
with $z_j=r_j e^{i\varphi_j}\in\mathbb{C}$, $j=1,\ldots,N$. The coupling matrix
$A=\{a_{ij}\}_{i,j=1}^N$ determines the topology of the network. The
local dynamics of each element is given by the normal form of a
supercritical Hopf bifurcation, also known as Stuart-Landau
oscillator,
\begin{equation}
    \label{HopfEq}
    f(z_j)=[\lambda + i\omega - (1 + i\gamma)|z_j|^2]z_j
\end{equation}
with real constants $\lambda, \omega\neq 0$, and $\gamma$. In
Eq. (\ref{mainEq}), $\tau$ is the delay time. $K$ and $\beta$ denote
the amplitude and phase of the complex coupling constant, respectively. Such kinds
of networks are used in different areas of nonlinear dynamics, e.g.,
to describe neural activities \cite{HAU07a}.
% \cite{Haup07}.

Synchronous in-phase, cluster, and splay states are possible solutions
of Eqs. (\ref{mainEq}) and (\ref{HopfEq}). They exhibit a common 
amplitude $r_j\equiv r_{0,m}$ and phases given by $\varphi_j=\Omega_m
t+j\Delta\varphi_m$ with a phase shift $\Delta\varphi_m=2\pi m/N$
and collective frequency $\Omega_m$. The
integer $m$ determines the specific state: in-phase oscillations
correspond to $m=0$, while splay and cluster states correspond to
$m=1,\dots,N-1$.
The cluster number $d$, which determines how many clusters of oscillators exist, 
is given by the least common multiple of $m$ and $N$ divided by $m$, and
$d=N$ (e.g., $m=1$), corresponds to a splay state. 

The stability of synchronized oscillations in networks can be
determined numerically, for instance, by the \textit{master stability
function}~\cite{PEC98}. This formalism allows a separation of the
local dynamics of the individual nodes from the network topology. In
the case of the Stuart-Landau oscillators it was possible to obtain
the Floquet exponents of different cluster states analytically with
this technique \cite{CHO09}. By these means it has been demonstrated
that the unidirectional ring configuration of Stuart-Landau
oscillators exhibits in-phase synchrony, splay states, and clustering
depending on the choice of the control parameter $\beta$. For
$\beta=0$, there exists multistability of the possible synchronous states in a
large parameter range. However, when tuning the coupling phase to an
optimal value $\beta=\Omega_m\tau-2\pi m/N$ according to a particular
state $m$, this synchronous state is monostable for any values of the
coupling strength $K$ and the time delay $\tau$.  The main goal of this paper is
to find adequate values of $\beta$ by automatic adaptive adjustment.  For this
purpose, we make use of the speed gradient method \cite{FRA07}, which
is outlined in the next section.

\section{Speed-gradient method}\label{sec:SG}

In this section, we briefly review an adaptive control scheme called 
\textit{speed-gradient} (SG) \textit{method}.
Consider a general nonlinear dynamical system
\begin{equation}
    \label{SG_sys}
    \dot x = F(x, u, t)
\end{equation}
with state vector $x\in\mathbb{C}^n$, input (control) variables $u\in\mathbb{C}^m$,
and nonlinear function $F$. Define a control goal 
\begin{equation}
\lim_{t\to\infty} Q(x(t), t) = 0,
\end{equation}
where $Q(x,t) \ge 0$ is a smooth scalar goal function.

In order to design a control algorithm, the scalar function
$\dot{Q}=\omega(x,u,t)$ is calculated, that is, the speed (rate) at
which $Q(x(t),t)$ is changing along trajectories of Eq.~(\ref{SG_sys}): 
\begin{equation}
\omega(x,u,t)=\frac{\partial Q(x,t)}{\partial t}+[\nabla_x Q(x,t)]\trn F(x,u,t).
\end{equation}
Then we evaluate the gradient of
$\omega(x,u,t)$ with respect to input variables: 
%$\nabla_u \omega(x,u,t)=(\frac{\partial\omega}{\partial u})\trn=(\frac{\partial
%  F}{\partial u})\trn\nabla_x Q(x,t)$. 
$$\nabla_u \omega(x,u,t)=\nabla_u [\nabla_x Q(x,t)]\trn F(x,u,t).$$ 
Finally, we set up a
differential equation for the input variables $u$
\begin{equation}
\label{SGAlgorithm} \frac{du}{dt}=-\Gamma\nabla_u\omega(x,u,t),
\end{equation}
where $\Gamma=\Gamma\trn>0$ is a positive definite gain matrix.
% , e.g., $\Gamma=\operatorname{diag}\{\gamma_1,\ldots,\gamma_m\}$, $\gamma_i>0$.
The algorithm (\ref{SGAlgorithm}) is called \textit{speed-gradient}
(SG) \textit{algorithm}, since it suggests to change $u$
proportionally to the gradient of the speed of changing $Q$.

The idea of this algorithm is the following. The term $-\nabla_u\omega(x,u,t)$ 
points to the direction in which the value of $\dot Q$ 
decreases with the highest speed. 
Therefore, if one forces the control signal to "follow" this direction,
the value of $\dot Q$ 
will decrease and finally be negative. When $\dot Q <0$, 
then $Q$ will decrease and, eventually, tend to zero.

We shall now apply the speed-gradient method to networks of
Stuart-Landau oscillators. Since the coupling phase $\beta$ is the
crucial parameter that determines stability of the possible in-phase,
cluster, and splay states,
we use this control parameter as the input variable $u$.
Setting $u=\beta$ and $x=(z_1,\ldots,z_N)$, Eq.~\eqref{mainEq} takes the form
of Eq.(\ref{SG_sys})  
with state vector $x\in\mathbb{C}^N$ and input variable
$\beta\in\mathbb{R}$, and nonlinear function
$F(x,\beta,t)=[f(z_1),\ldots,f(z_N)] + Ke^{i\beta} [ A x(t-\tau) -
x(t)]$.

The SG control equation (\ref{SGAlgorithm}) for the input
variable $\beta$ then becomes
 \begin{equation}
  \label{SGAlgorithm_beta}
  \frac{d\beta}{dt}=-\Gamma \frac{\partial}{\partial \beta}\omega(x,\beta,t)
=-\Gamma \left(\frac{\partial F}{\partial \beta} \right)\trn\nabla_x Q(x,t),
\end{equation}
where $\Gamma>0$ is now a scalar.

\section{In-phase synchronization}
\label{sec:in-phase-synchr}

To apply the SG method for the selection of in-phase synchronization
we need to find an appropriate goal function $Q$. It should satisfy the
following conditions: the goal function must be zero for an in-phase
synchronous state and larger than zero for other states. Hence, a
simple goal function can be introduced by taking the distance of all
oscillator phases to a reference oscillator's phase $\varphi_1$:
\begin{equation}
\label{goalPhase1} Q_1(x(t),t)=\frac{1}{2}\sum_{k=2}^N(\varphi_k-\varphi_1)^2,
\end{equation}
%where $z_j=r_je^{i\varphi_j}$. 
Taking the gradient of the derivative
along the trajectories of the system (\ref{mainEq}) with local
dynamics (\ref{HopfEq}) one can derive an adaptive law of the
following form by straight-forward calculation. Using
$\omega(x,\beta,t)=\dot{Q}_1$, Eq.~\eqref{SGAlgorithm_beta} becomes
\begin{widetext}
\begin{equation}
  \label{BetaEq}
  \dot{\beta}=-\Gamma K\sum_{k=2}^N(\varphi_k-\varphi_1)\left[\sum_{n=1}^N
    a_{kn}\left(\frac{r_{n,\tau}}{r_k}\cos(\beta+\varphi_{n,\tau}
      -\varphi_k)-\cos\beta\right)-
    \sum_{n=1}^N
    a_{1n}\left(\frac{r_{n,\tau}}{r_1}\cos(\beta+\varphi_{n,\tau}
      -\varphi_1)-\cos\beta\right)\right],
\end{equation}
\end{widetext}
where we used the abbreviations $r_{n,\tau}=r_n(t-\tau)$ and
$\varphi_{n,\tau}=\varphi_n(t-\tau)$ for notational convenience.

\begin{figure}[th!]
\includegraphics[width=\linewidth]{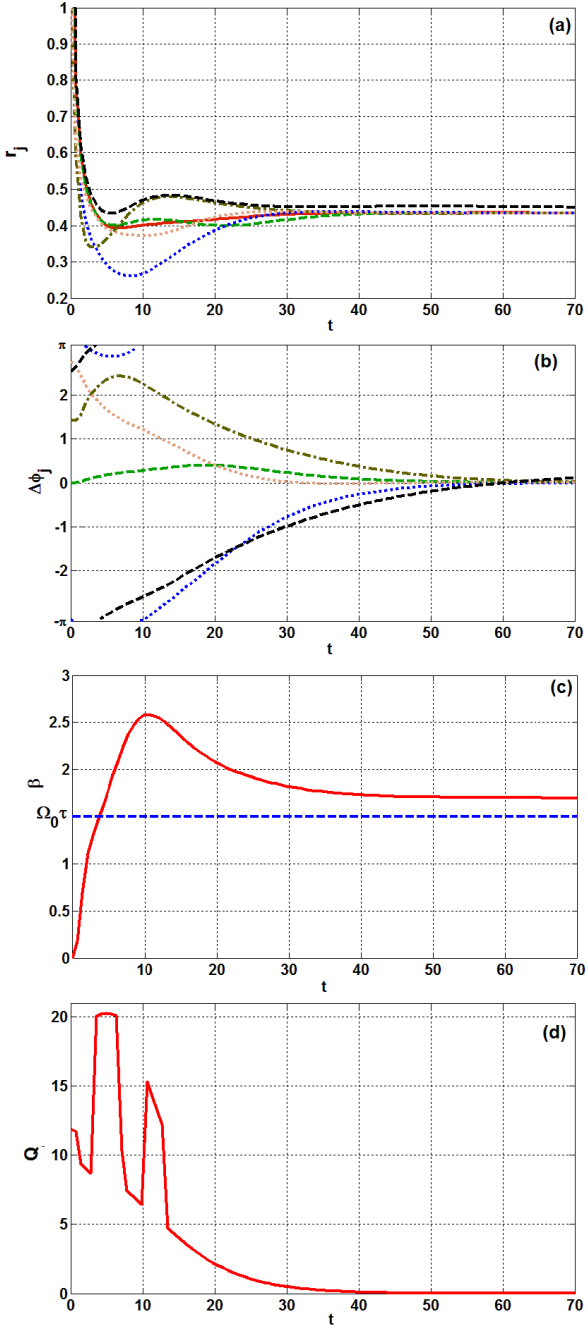}
\caption{(Color online) Adaptive control of in-phase oscillations 
with goal function Eq.~(\ref{goalPhase1}).
(a): absolute values $r_j=\left|z_j\right|$ for $j=1,...,6$; (b):
  phase differences $\Delta\phi_j=\varphi_j-\varphi_{j+1}$ for 
$j=1,...,5$; 
  (c) temporal evolution of $\beta$, blue dashed line: reference value 
  for $\Omega_0=0.92$; (d): goal function. Parameters: 
$\lambda=0.1, \omega=1, \gamma=0$, $K=0.08$, $\tau=0.52 \pi$, $N=6$.
Initial conditions for $r_j$ and $\varphi_j$ are chosen randomly from
$[0,4]$ and $[0, 2\pi]$, respectively. The initial condition for $\beta$ is zero.}
\label{in-phase1}
\end{figure}

Figure~\ref{in-phase1} presents the results of a numerical simulation
for a random network with $N=6$ nodes and unity row sum. 
Throughout this paper we use $\Gamma=1$. According to the
numerical simulations decreasing $\Gamma$ will yield a decrease of the
speed of convergence. On the other hand, if $\Gamma$ is too big,
undesirable oscillations appear.  The model parameters are
chosen as in \cite{CHO09}. %$\lambda=0.1, \omega=1, \gamma=0$. 
In Fig.~\ref{in-phase1}(a) it can be seen that the absolute values $|z_j|$ of all 
nodes converge after about 60 time units. Fig.~\ref{in-phase1}(b) 
shows that the phase differences of the different oscillators approach zero, 
which corresponds to the
in-phase synchronous state. Fig.~\ref{in-phase1}(c) depicts the evolution of
$\beta$. The blue dashed line represents the value of the coupling phase 
$\beta=\Omega_0\tau=0.48\pi$, for which stability was shown analytically in \cite{CHO09}.
It can be seen that the adaptively 
adjusted phase comes close to this value. In other words, even without knowing the
exact values of the system parameters, the SG algorithm yields an
adequate value of $\beta$ that stabilizes the target state of in-phase
synchronization. Fig.~\ref{in-phase1}(d) shows that the
goal function (\ref{goalPhase1}) indeed approaches zero.

\begin{figure}[th!]
\includegraphics[width=\linewidth]{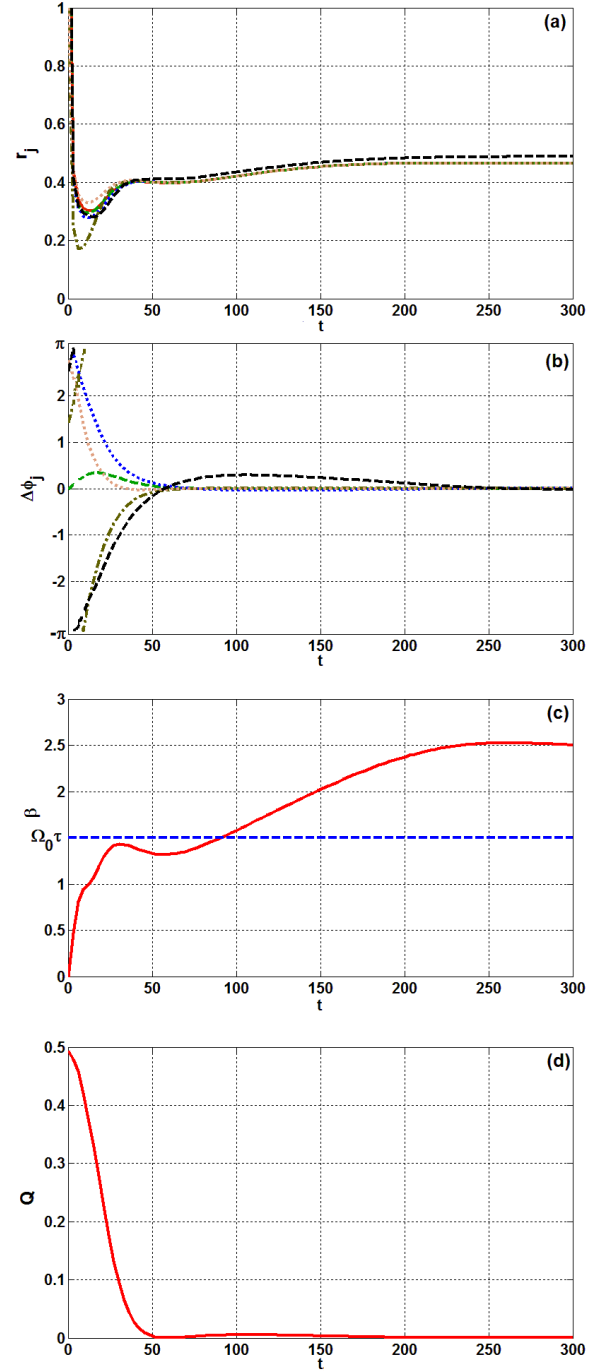}
\caption{(Color online) Adaptive control of in-phase oscillations 
with goal function Eq.~(\ref{goalPhase2}). (a): absolute values $r_j=\left|z_j\right|$; 
(b): phase differences $\Delta\phi_j=\varphi_j-\varphi_{j+1}$; 
(c): temporal evolution of $\beta$, blue dashed line: reference value for $\Omega_0=0.92$;
(d): goal function.
Other parameters as in Fig.~\ref{in-phase1}.}
\label{in-phase2}
\end{figure}

Note that the above choice of the goal function $Q$ is not the only possibility to 
generate a stable in-phase solution. Let us consider a function based on the order 
parameter
\begin{equation}
    R_1=\frac{1}{N}\left|\sum_{j=1}^N e^{i\varphi_j}\right|.
\end{equation}
It is obvious that $R_1=1$ if and only if the state is in-phase synchronized. For
other cases we have $R_1<1$. Using this observation we can introduce the following 
goal function
\begin{equation}
    Q_2=1-\frac{1}{N^2}\sum_{j=1}^N e^{i\varphi_j}\sum_{k=1}^N e^{-i\varphi_k}.
    \label{goalPhase2}
\end{equation}

From $\dot{\beta}=-\Gamma \frac{\partial}{\partial \beta} \dot{Q}_2$ we derive an
alternative adaptive law:
\begin{widetext}
\begin{equation}
  \label{BetaEqq}
  \dot{\beta}=\Gamma\frac{2K}{N^2}\sum_{k=1}^N\sum_{j=1}^N \sin(\varphi_k-\varphi_j)\sum_{n=1}^N a_{jn} (\frac{r_{n,\tau}}{r_j}\cos(\beta+\varphi_{n,\tau}-\varphi_j)-\cos\beta). 
\end{equation}
\end{widetext} 
Fig.~\ref{in-phase2} shows the
results of a numerical simulation. As before, the amplitude and phase
approach appropriate values that lead to in-phase synchronization.
This time, however, the obtained value of $\beta$ does not converge to the one for which
the analytical approach \cite{SCH09} has established stability of the in-phase 
oscillation (blue dashed line), but to another limit value.
This can be explained as follows: There exists a whole interval of 
acceptable values of $\beta$ around
the value of the coupling phase for which an analytical treatment is possible,
such that for any value from this interval an in-phase state is stable. Our
SG algorithm finds one of them, depending upon initial conditions.
%  but not necessary the one that was obtained theoretically.

\section{Splay and cluster states stabilization}
\label{sec:splay-cluster-states}

In this section we will consider unidirectionally coupled rings 
with $N=6$ nodes. That is, the coupling matrix has the following form:
$$
    A=\begin{pmatrix}
        0 & 1 & 0 &\cdots & 0\\
        0 & 0 & 1 &\cdots & 0\\
        \vdots& \vdots& \vdots & \ddots & \vdots\\
        0 & 0 & 0 &\cdots & 1\\
        1 & 0 & 0 &\cdots & 0
    \end{pmatrix}
$$
Let $1\le m\le N-1$. Then $d=\operatorname{LCM}(m,N)/m$, where LCM
denotes the least common multiple, is the number of different
clusters of a synchronized solution. A splay state corresponds to
$d=N$ while cluster states yield $d<N$. Using similar arguments as those
leading to Eq.~(\ref{goalPhase1}) we could choose a goal
function of the following form:
\begin{equation}
    \label{goalPhase3}
Q_3=\frac{1}{2}\sum_{j=1}^N\left(\varphi_j-\varphi_{j+1}-\frac{2\pi}{d}
\right)^2
\end{equation}
with $j=j \mod N$.
% where $\varphi_{N+1}\triangleq\varphi_1. $

\begin{figure}[th!]
\includegraphics[width=\linewidth]{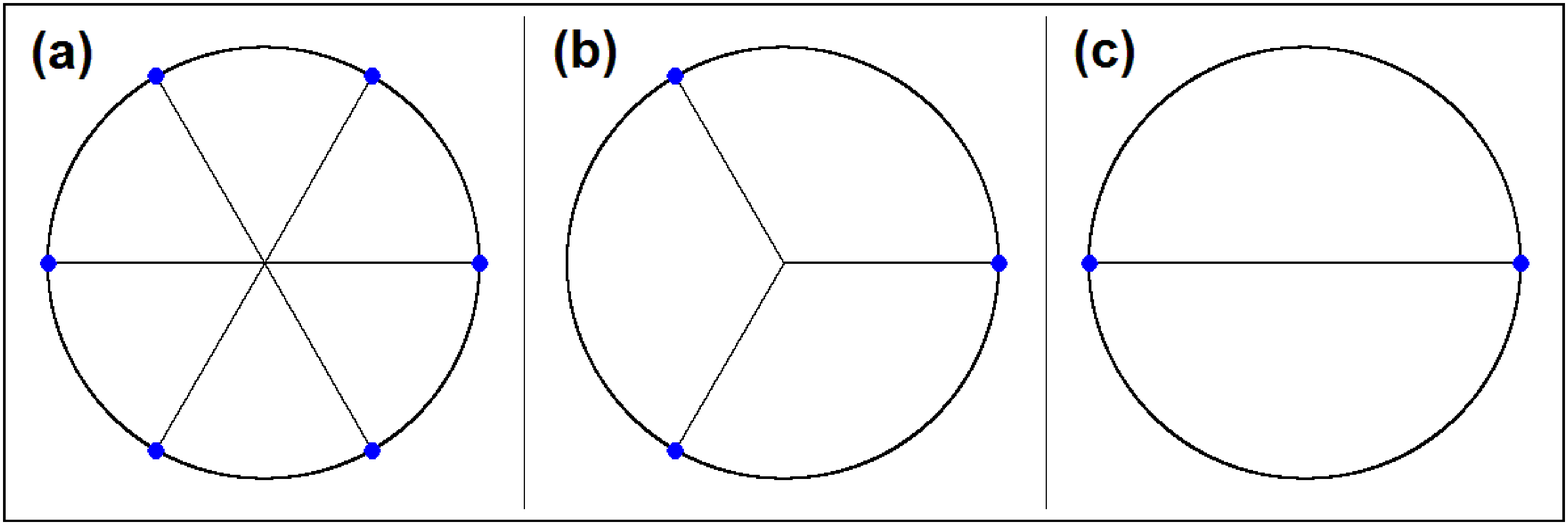}
\caption{Schematic diagrams of splay ($d=6$), three-cluster ($d=3$), and two-cluster
($d=2$)  states in panels (a), (b), and (c), respectively ($N=6$).
Each cluster contains the same number of nodes.}
\label{notes}
\end{figure}

The goal function Eq.~(\ref{goalPhase3}) has a crucial disadvantage: we
need to define
% neighbour for each node, that is, we determine
an ordering of the system nodes. Since this is inconvenient for
practical applications, we will extend the alternative goal function
Eq.~(\ref{goalPhase2}) such that we can stabilize splay and cluster states. 
First of all, note that the following
condition holds for splay and cluster states:
\begin{equation}
\label{note1}
    \sum_{j=1}^Ne^{i\varphi_j}=0.
\end{equation}
Indeed, if we have only three nodes and take $Q=\sum_{j=1}^3
e^{i\varphi_j}\sum_{k=1}^3 e^{-i\varphi_k}$ as a goal function, we
will ensure stability of a splay state, as we have verified by
numerical simulations. Note that this goal function does not need a 
fixed ordering of the nodes. Renumbering all nodes in a
random way will yield the same goal function.  One can define
a generalized order parameter
\begin{equation}
\label{note2}
    R_d=\frac{1}{N}\left|\sum_{k=1}^N e^{d i\varphi_k}\right|
\end{equation}
with $d\in\mathbb{N}$.
However, if we derive a goal function from this order parameter 
in an analogous way as in Eq.~(\ref{goalPhase2}), this function will not have 
a unique minimum at the $d$-cluster state because $R_d=1$ holds also for the 
in-phase state and for other $p$-cluster states where $p$ are divisors of $d$.

For example, suppose that the system has six nodes. Then states for which
conditions (\ref{note1}) and (\ref{note2}) with $R_d=1$ for 
$d=6$ hold are schematically
depicted in Fig.~\ref{notes}(a),(b),(c). In order to distinguish between
these three cases, let us consider the functions
\begin{equation}
f_p(\varphi)=\frac{1}{N^2}\sum_{j=1}^N e^{pi\varphi_j}\sum_{k=1}^N
e^{-pi\varphi_k}. 
\label{def_f_p}
\end{equation}
A splay state (Fig.~\ref{notes}(a)) yields $f_1=f_2=f_3=0$, while in the 3-cluster state
displayed in Fig.~\ref{notes}(b) we have $f_1=f_2=0$, $f_3=1$, 
and in the 2-cluster-state shown in Fig.~\ref{notes}(c)
$f_1=f_3=0$, $f_2=1$. Hence, we obtain $\sum_p f_p=0$ if and only if
there is a state with $d$ clusters, where the sum is taken over all
divisors of $d$. x,t)

Combining all previous results we adopt the following goal function:
\begin{equation}
  Q_4=1-f_d(\varphi)+\frac{N^2}{2}\sum_{p|d, 1\le p<d}f_p(\varphi),
\label{goalPhase4}
\end{equation}
where $p|d$ means that $p$ is a factor of $d$.  This goal function contains
$f_d$ as the primary contribution for the $d$-cluster state, but also
a sum of penalty terms that counteract reaching other cluster states
in which $f_d$ is also unity. Whenever one of those unwanted cluster
states is approached, the penalty term will lead to a gradient away
from it. The prefactor $N^2/2$ is chosen for convenience to secure faster
convergence of the algorithm. 
From $\dot{\beta}=-\Gamma \frac{\partial}{\partial \beta}\dot{Q}_4$ 
one can derive the adaptation law
\begin{widetext}
\begin{equation}
  \dot{\beta}=-\Gamma K\sum_{j=1}^N\sum_{k=1}^N\left\{\sum_{p|d, 1\le
      p<d}p\sin[p(\varphi_k-\varphi_j)]-\frac{2d}{N^2}
    \sin[d(\varphi_k-\varphi_j)]\right\}\sum_{n=1}^N
  a_{jn}\left[\frac{r_{n,\tau}}{r_j}\cos(\beta+\varphi_{n,\tau}
    -\varphi_j)-\cos(\beta)\right].
\label{BetaEq4}
\end{equation}
\end{widetext}

\begin{figure}[th!]
\includegraphics[width=\linewidth]{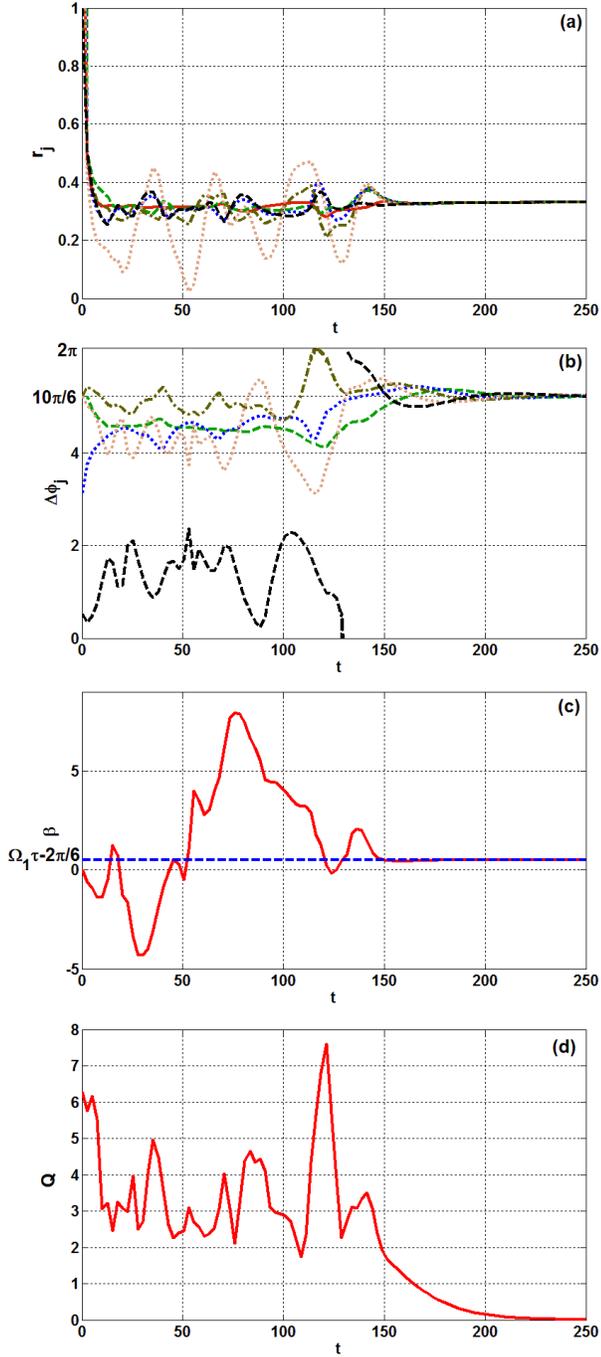}
\caption{(Color online) Adaptive control of splay state 
with goal function Eq.~(\ref{goalPhase4}). (a): absolute values $r_j=\left|z_j\right|$; 
(b): phase differences $\Delta\phi_j=\varphi_j-\varphi_{j+1}$; (c): 
temporal evolution of $\beta$, blue dashed line: reference value for $\Omega_1=0.96$; (d): goal function.
Other parameters as in Fig.~\ref{in-phase1}.}
\label{splay}
\end{figure}

\begin{figure}[th!]
\includegraphics[width=\linewidth]{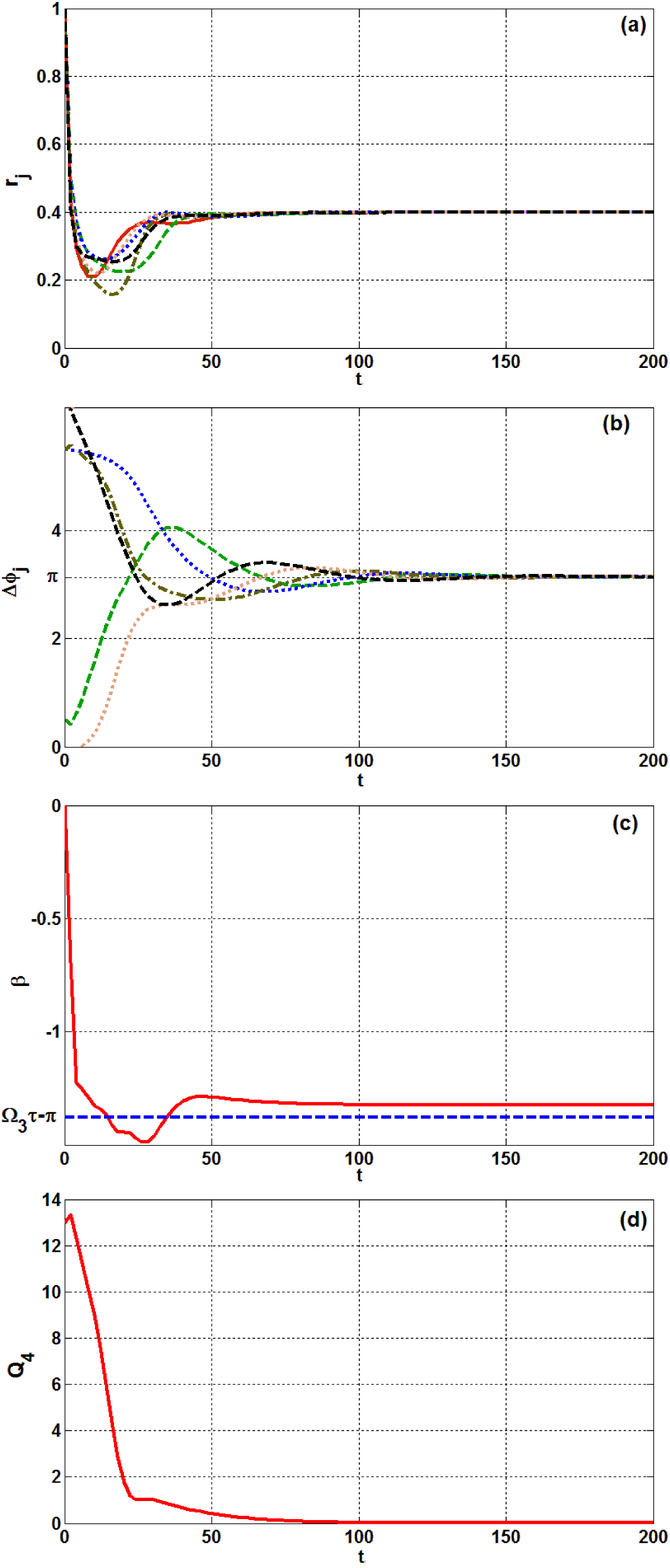}
\caption{(Color online) Adaptive control of 2-cluster state ($m=3$) 
with goal function Eq.~(\ref{goalPhase4}). (a): absolute values $r_j=\left|z_j\right|$; 
(b): phase differences $\Delta\phi_j=\varphi_j-\varphi_{j+1}$; (c): 
temporal evolution of $\beta$, blue dashed line: reference value for $\Omega_3=1.08$; (d): goal function.
Other parameters as in Fig.~\ref{in-phase1}.}
\label{clusterm3}
\end{figure}

\begin{figure}[th!]
\includegraphics[width=\linewidth]{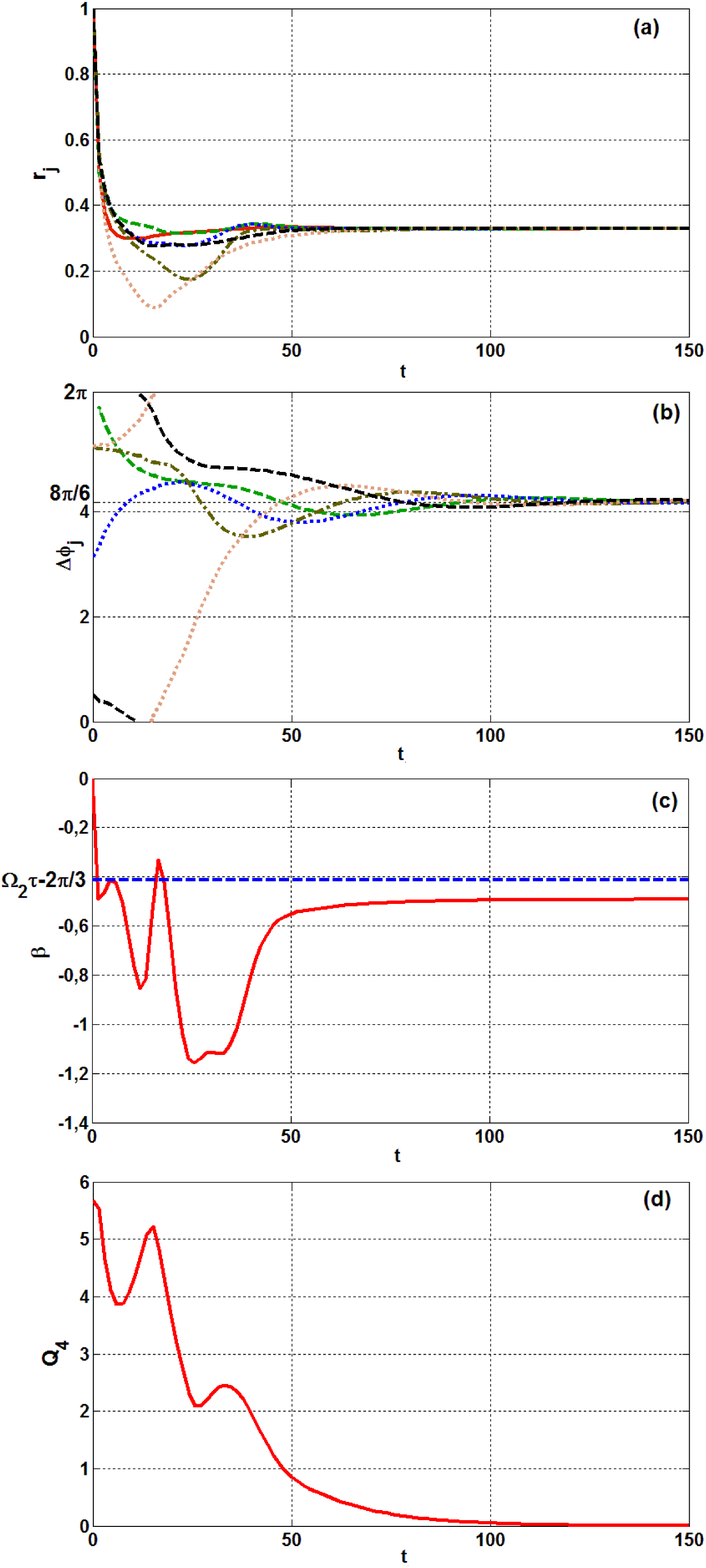}
\caption{(Color online) Adaptive control of 3-cluster state ($m=2,4$) 
with goal function Eq.~(\ref{goalPhase4}). (a): absolute values $r_j=\left|z_j\right|$; 
(b): phase differences $\Delta\phi_j=\varphi_j-\varphi_{j+1}$, 
blue dashed line: reference value for $\Omega_2=1.03$; (c): 
temporal evolution of $\beta$; (d): goal function.
Other parameters as in Fig.~\ref{in-phase1}.}
\label{clusterm2}
\end{figure}

In Fig.~\ref{splay} we show the results of a numerical simulation for
splay state stabilization ($d=N=6$, $m=1$). The phase differences are
$\Delta\phi_j=\varphi_j-\varphi_{j+1}=2\pi-2\pi/N$, which corresponds 
to the splay state. In
Fig.~\ref{splay}(c) one can see that the adaptively obtained value of
$\beta$ converges to that for which stability was shown analytically in \cite{CHO09}
(dashed blue line).

Figures~\ref{clusterm3} and \ref{clusterm2} depict the results of
numerical simulations for two clusters ($d=2$, $m=3$) and
three clusters ($d=3$, $m=2,4$), respectively. Again
we note that the obtained value of $\beta$ comes close to the one
for which stability was shown analytically in \cite{CHO09}.

The above results indicate that the speed-gradient method is able to drive
the network dynamics into the desired cluster or splay state by
adaptively adjusting the coupling phase, where the goal function is
chosen according to the corresponding target state. We have, however, used
only exemplary values of the coupling parameters $K$ and $\tau$ so
far.

\begin{figure}[th!]
  \centering
  \includegraphics[width=\linewidth]{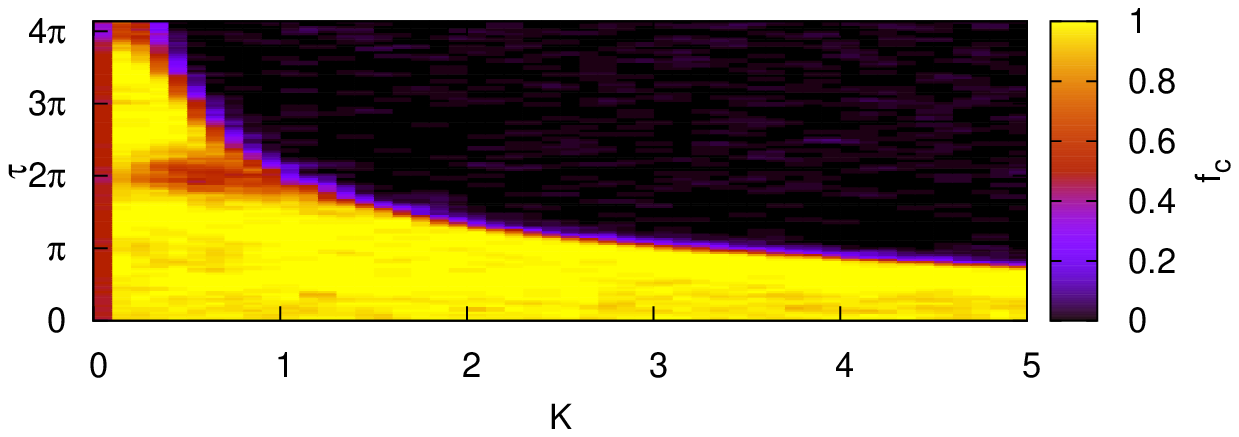}
  \caption{(Color online) Success of the speed-gradient method in dependence on the
    coupling parameters $K$ and $\tau$ for the splay state in a
    unidirectionally coupled ring of $N=4$ Stuart-Landau oscillators. 
    Other parameters as in Fig.~\ref{in-phase1}.
    The color code shows the fraction of successful realizations. }
  \label{fig:K-tau-plane}
\end{figure}
For the example of a splay state (4-cluster) in a network of 
4 Stuart-Landau oscillators coupled in a unidirectional ring
we have conducted a more exhaustive analysis of the ($K,\tau$) plane. 
Figure~\ref{fig:K-tau-plane} shows
results in dependence on the coupling strength $K$ and the coupling
delay $\tau$. According to Ref.~\cite{CHO09} there exists an optimal
value of the coupling phase that enables stability of this state for
arbitrary values of $K$ and $\tau$. We ran simulations with 20
different initial conditions chosen randomly from the complex interval
$[-1,1] \times [-i,i]$ for each oscillator $z_j$. Figure~\ref{fig:K-tau-plane} 
shows the fraction $f_c$ of those realizations that asymptotically approach 
a splay state after applying the speed-gradient method. We observe that 
the speed-gradient method is able to control the splay state in a wide 
parameter range. The range of possible coupling strengths $K$ does, however, 
shrink considerably with increasing time delay $\tau$. We conjecture several 
reasons for this shrinking. Firstly, multistability of different splay and cluster
states is more likely for larger values of $K$ and $\tau$, which
narrows down the basin of attraction for a given state. Secondly,
Eq.~\eqref{BetaEq4}, 
which describes the dynamics of the coupling phase
under the adaptive control, is influenced by the time delay
$\tau$. Using large delay times, we observe overshoots of the control
leading to a failure. 

\section{Conclusion}\label{sec:conclusions}
We have proposed a novel adaptive method for the control of 
synchrony on oscillator networks,
which combines time-delayed coupling with the speed gradient 
method of control theory. Choosing an appropriate goal function,
a desired state of generalized synchrony can be selected
by the self-adaptive automatic adjustment of a
control parameter, i.e., the coupling phase. This goal function, which
is based on a generalization of the Kuramoto order parameter, vanishes
for the desired state, e.g., in-phase, splay, or cluster states,
irrespectively of the ordering of the nodes. By numerical simulations we
have shown that those different states
can be stabilized, and the coupling phase
converges to an optimum value. We have elaborated on
the robustness of the control scheme by investigating the success rates 
of the algorithm in dependence on the coupling parameters, i.e., the coupling strength
and the time delay. In this work, we focused on the
adaptive adjustment of the coupling phase while the other coupling
parameters were fixed. The input variable $u$ in Eq.~\eqref{SG_sys} may in
general contain all of the coupling parameters. Thus, as a generalization, our method might be applied
to all coupling parameters including the coupling amplitude 
and the time delay.  In this way control of cluster and splay
synchronization might be possible without any a priori knowledge of the coupling
parameters. Given the paradigmatic nature of the Stuart-Landau
oscillator as a generic model, we expect broad applicability, for instance 
to synchronization of networks in medicine, chemistry or mechanical
engineering. The mean-field nature of our goal function makes our
approach accessible even for very large networks independently of the
particular topology.

\begin{acknowledgments}
  This work is supported by the German-Russian Interdisciplinary
  Science Center (G-RISC) funded by the German Federal Foreign Office
  via the German Academic Exchange Service (DAAD). JL, TD, PH, and ES
  acknowledge support by Deutsche Forschungsgemeinschaft (DFG) in the
  framework of SFB 910.
\end{acknowledgments}

% \bibliographystyle{apsrev4-1}
% \bibliography{bibliography}
%\bibliographystyle{prsty-fullauthor}
%\bibliography{ref}

\begin{thebibliography}{10}

\bibitem{SCH07}
{\em Handbook of Chaos Control}, edited by E. Sch{\"o}ll and H.~G. Schuster
  (Wiley-VCH, Weinheim, 2008), second completely revised and enlarged edition.

\bibitem{PYR92}
K. Pyragas, Phys.~Lett.~A {\bf 170},  421  (1992).

\bibitem{SOC94}
J.~E.~S. Socolar, D.~W. Sukow, and D.~J. Gauthier, Phys.~Rev.~E {\bf 50},  3245
   (1994).

\bibitem{PYR06}
V. Pyragas and K. Pyragas, Phys.~Rev.~E {\bf 73},  036215  (2006).

\bibitem{POP06}
O.~V. Popovych, C. Hauptmann, and P.~A. Tass, Biol. Cybern. {\bf 95},  69
  (2006).

\bibitem{SCH06a}
S. Schikora, P. H{\"o}vel, H.~J. W{\"u}nsche, E. Sch{\"o}ll, and F.
  Henneberger, Phys.~Rev.~Lett. {\bf 97},  213902  (2006).

\bibitem{ZHA08}
Y. Zhai, I.~Z. Kiss, and J.~L. Hudson, Ind.~Eng.~Chem.~Res. {\bf 47},  3502
  (2008).

\bibitem{ORO10}
G. Orosz, J. Moehlis, and R.~M. Murray, Phil. Trans.~R. Soc.~A {\bf 368},  439
  (2010).

\bibitem{DAH10}
T. Dahms, V. Flunkert, F. Henneberger, P. H{\"o}vel, S. Schikora, E.
  Sch{\"o}ll, and H.~J. W{\"u}nsche, Eur. Phys.~J.~ST {\bf 191},  71  (2010).

\bibitem{SCH09}
E. Sch{\"o}ll,  in {\em Nonlinear Dynamics of Nanosystems}, edited by G.
  Radons, B. Rumpf, and H.~G. Schuster (Wiley-VCH, Weinheim, 2010), pp.\
  325--367.

\bibitem{SCH09a}
E. Sch{\"o}ll, P. H{\"o}vel, V. Flunkert, and M.~A. Dahlem,  in {\em {Complex
  time-delay systems: theory and applications}}, edited by F.~M. Atay
  (Springer, Berlin, 2010), pp.\ 85--150.

\bibitem{FLU11}
V. Flunkert and E. Sch{\"o}ll, Phys. Rev. E  (2011), in print.

\bibitem{FRA79}
A.~L. Fradkov, Autom. Remote Control {\bf 40},  1333  (1979).

\bibitem{FRA05b}
A.~L. Fradkov, Physics-Uspekhi {\bf 48},  103  (2005).

\bibitem{FRA07}
A.~L. Fradkov, {\em {Cybernetical Physics: From Control of Chaos to Quantum
  Control}} (Springer, Heidelberg, Germany, 2007).

\bibitem{KEH09}
M. Kehrt, P. H{\"o}vel, V. Flunkert, M.~A. Dahlem, P. Rodin, and E. Sch{\"o}ll,
  Eur. Phys. J. B {\bf 68},  557  (2009).

\bibitem{HOE09}
P. H{\"o}vel, M.~A. Dahlem, and E. Sch{\"o}ll, Int.~J.~Bifur.~Chaos {\bf 20},
  813  (2010).

\bibitem{FLU10b}
V. Flunkert, S. Yanchuk, T. Dahms, and E. Sch{\"o}ll, Phys.~Rev.~Lett. {\bf
  105},  254101  (2010).

\bibitem{OME11}
I. Omelchenko, Y. Maistrenko, P. H{\"o}vel, and E. Sch{\"o}ll, Phys. Rev. Lett.
  {\bf 106},  234102  (2011).

\bibitem{BRO11}
G. Brown, C.~M. Postlethwaite, and M. Silber, Physica~D {\bf 240},  859
  (2011).

\bibitem{CHO09}
C.~U. Choe, T. Dahms, P. H{\"o}vel, and E. Sch{\"o}ll, Phys. Rev.~E {\bf 81},
  025205(R)  (2010).

\bibitem{CHO11}
C.~U. Choe, T. Dahms, P. H{\"o}vel, and E. Sch{\"o}ll,  in {\em Proceedings of
  the Eighth AIMS International Conference on Dynamical Systems, Differential
  Equations and Applications} ({American Institute of Mathematical Sciences},
  Springfield, MO, USA, 2011), in print.

\bibitem{FLU07}
V. Flunkert and E. Sch{\"o}ll, Phys. Rev. E {\bf 76},  066202  (2007).

\bibitem{FIE07}
B. Fiedler, V. Flunkert, M. Georgi, P. H{\"o}vel, and E. Sch{\"o}ll,
  Phys.~Rev.~Lett. {\bf 98},  114101  (2007).

\bibitem{SCH11}
S. Schikora, H.~J. W{\"u}nsche, and F. Henneberger, Phys. Rev.~E {\bf 83},
  026203  (2011).

\bibitem{PYR08}
K. Pyragas and T. Pyragiene, Phys. Rev. E {\bf 78},  046217  (2008).

\bibitem{HAU07a}
C. Hauptmann, O. Omel`chenko, O.~V. Popovych, Y. Maistrenko, and P.~A. Tass,
  Phys. Rev. E {\bf 76},  066209  (2007).

\bibitem{PEC98}
L.~M. Pecora and T.~L. Carroll, Phys. Rev. Lett. {\bf 80},  2109  (1998).

\end{thebibliography}

\end{document}